# DEVELOPMENT OF SECURITY DETECTION MODEL FOR THE SECURITY OF SOCIAL BLOGS AND CHATTING FROM HOSTILE USERS

Shubhankar Gupta[1] and Nitin[2]

[1]Jaypee Institute of Information Technology, Noida, India

[2]Professor, CEAS, University of Cincinnati, Cincinnati, USA

## ABSTRACT

*Worldwide, a large number of people interact with each other by means of online chatting. There has been a significant rise in the number of platforms, both social and professional, such as WhatsApp, Facebook, and Twitter, which allow people to share their experiences, views and knowledge with others. Sadly enough, with online communication getting embedded into our daily communication, incivility and misbehaviour has taken on many nuances from professional misbehaviour to professional decay. Generally flaming starts with the exchange of rude messages and comments, which in turn triggers to higher scale of flaming. To prevent online communication from getting downgraded, it is essential to keep away the hostile users from communication platforms. This paper presents a Security Detection Model and a tool which checks and prevents online flaming. It detects the presence of flaming while chatting or posting blogs, and censors swear words as well as blocks the users from flaming.*

## 1. INTRODUCTION

In the last few years, online communication has increased manifold with millions of people discovering the power of Information Technology (IT) each month. Throughout the world, a large number of people interact with each other through the means of online chatting which has now overtaken other forms of communication. Recently, there has been a significant rise in the number of platforms, both social and professional, such as WhatsApp, Facebook, Twitter, Quora, which allow people to share their experiences, views, knowledge, and business communications with others [1]. Nowadays, online forums and social networking sites are becoming a platform for business and socially active people to engage with each other for business networking, discussing issues and promoting businesses by sharing online description of products, tools or solutions to different kinds of problems.

Undoubtedly, there are several advantages of the mode of online communication, but there are some disadvantages attached to it in the form of online aggressive behaviour, known as *flaming* in IT industry. The term flaming originated from the *Hacker's Dictionary* [2], which defines it as to *speak rapidly or incessantly on an uninteresting topic or with a patently ridiculous attitude*. The meaning of the word has diverged from this definition since then. In fact, *flaming is a hostile and insulting interaction between persons, often involving the use of profanity*. It can also be the swapping of insults back and forth or with many people in a group or on a single person. It can be deliberately provoked by seemingly trivial differences. Deliberate flaming, as opposed to flaming as a result of emotional discussions, is carried out by individuals known as flamers or troller or bullies, who are specifically motivated to incite flaming. According to [3], distinct messages that are precipitative, often personally derogatory, ad hominem attacks directed toward someone due to a position taken in a message distributed to the group. Further, flaming is generally classified into numerous categories such as direct flaming, indirect flaming, straight flaming and satirical flaming. However, for this research work, we have predefined flaming only in four sub-classes as *hostile, aggressive, offensive and uninhibited*. For the purpose of measuring the resulting intensity, the ten increments on the scale are assigned scores of one to ten, with one indicating the lowest intensity and ten indicating the highest flaming intensity.

Hence it has become very important to keep hostile users or antagonistic criticism away from the online platform to promote healthy and topic centric communication. In this paper, in section 2 we have discussed flaming in chatting applications and what is flame detector tool. Section 3 briefly discusses our research methodology. Section 4 describe in detail, how we have implemented a Security Detection Model and developed a tool that will prevent the exchange of censored words between users, track the flaming level of a user and block users who indulge in flaming during online communications. It also discuss the experimental results and limitations of our Security Detection Model. The paper demonstrates that our model is working and achieving the intended goals.

## 2. FLAMING IN CHATTING APPLICATIONS AND FLAME DETECTOR TOOL

Social networking sites are also used to contact colleagues, friends and relatives. However, with the advances in social and technological platforms, people have also started interacting with strangers through social networking sites [4, 5, 6, 7, 8, 9, 10, 11, 12, 13, 14]. Out of these, strangers' research has also shown that males have a greater tendency to flame than female participants [4]. Even in the male population, it is the young, immature minds, who would account for higher flaming.

As mentioned above, chatting applications are in greater use because of excellent user interface as well as due to their user-friendly nature. Almost each and every user is registered on at least one or more chatting applications. According to the survey conducted by the *theverge.com* [15], 80% of the users use 1 to 5 chatting applications, 17% of the users use 6 to 10 applications and only 1% use 11 or more applications simultaneously. Two forms of chatting take place, one-to-one and one-to-many (group chats). Whichever may be the form, there are certain conventions or guidelines that have been created to avoid misunderstandings and to simplify communication between the users which is often called as *chatiquette*. Most of the users have not adhered to the convention which leads to uncivilized communication. Users tend to send text messages which are interpreted to be rude or sometimes abusive in nature and hence they might end up in conflict. Sometimes, people knowingly misspell a censor word by adding additional character or deleting one or more character to avoid flame detection. Such kind of behaviour has been reported from *Arabic-language* tweets [16, 17]. Eventually such forms of communication degrade the sense of camaraderie among friends, business people and others.

It is essential that some kind of integrity is maintained among the individuals. This kind of communication is more common among the youth who seem to be very aggressive in nature. In the professional world, such exchange of messages is not admissible. The diplomatic infringe of chatiquette engrosses the use of flames. Moreover, the more the users indulge in the use of rigorous language, the higher the flaming level tends to get.

We have developed a tool that mainly detects the presence of flaming while chatting or while posting blogs, and censors the swear words as well as blocks the users whose flaming levels are high (on a scale of 1-10). When a client sends a message, first the message is filtered of censored words at the server and flaming level assigned to the user and thereafter the filtered message is broadcasted to the other clients. This tool can be used for professional interactions where integrity needs to be maintained among the users and can be widely used in future.

A Flame Detector Tool has been proposed to detect the occurrence of flaming in online communication. This tool is installed on the local system and it gets plugged into any of the social networking sites such as Facebook, Twitter, Quora and chatting applications such as WhatsApp, Telegram and others. This tool consists of three components as shown in Figure 1. It consists of Social Networking Sites and Chatting Applications, Web Services and Flame Detector.

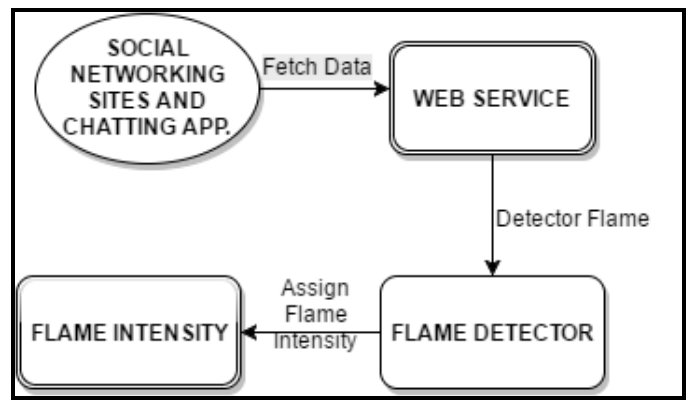

**Figure 1**. Flame Detector Schematic

As shown in figure 1, the flame detector tool has three components. The first component is social networking sites where formal as well as informal discussion takes place and the chatting applications from where the model gets its input. The second component is the Web Service which is used to fetch the data from social networking sites i.e. Facebook, Twitter which is then analysed by the Flame Detector to assign flame intensity to the user. We have developed a predefined thesaurus (censor.txt file), which contains words that are considered as flames. In addition to this, to avoid hostile users from flaming, there should be some security detection model that defends the networking sites and applications from flaming activities.

## 3. RESEARCH METHODOLOGY

Our research is centered upon the following research questions:

1. Is the Security Detection Model successful in blocking the users who indulge in flaming during chatting with others or while posting blogs using the Flame Detector Tool?
2. Is the Security Detection Model using the Flame Detector Tool successful in detecting the flames in the communications that takes place between the clients?

A research has been conducted to answer the above questions. In the next section, we have provided the architectural details of the tool based on Security Detection Model that we have described in this paper.

## 4. SECURITY DETECTION MODEL

The default flame value assigned to a user is zero if the latter is posting blog or sending message for the first time. Subsequently, if the user logs in ever again, the associated flaming level is retrieved. Over a period of time as the user exchanges data either in the form of text messages to others or in the form of blogs, the flaming level increases in the step unit of one, whenever certain amount of censored words are shared by the user. As long as the flaming level is below five, the user is not included in the category of hostile users and is given a warning. Flaming level above or equal to seven automatically blocks the user from sharing any kind of data for a specific period of time. The categories of flame earlier have been predefined here as hostile, aggressive, offensive and uninhibited. Hence this tool has been developed based on a Security Detection Model. The process of sending messages from client-to-client or to-many-clients using the Security Detection Model has been shown in figure 2

As shown in figure 2, when a client sends a text message, it is first passed to the server. The server then performs some operations by sending the received text message for scanning of censor words, if any present. The list of censor words is stored in a text file which is managed by the administrator. If any of the words mentioned in the message is present in the list, then it is replaced by an asterisk (*) and flaming intensity assigned to the user is incremented. In some cases, if the administrator finds any words in the message which are inappropriate, it can dynamically be added to the list of censor words for future convenience.

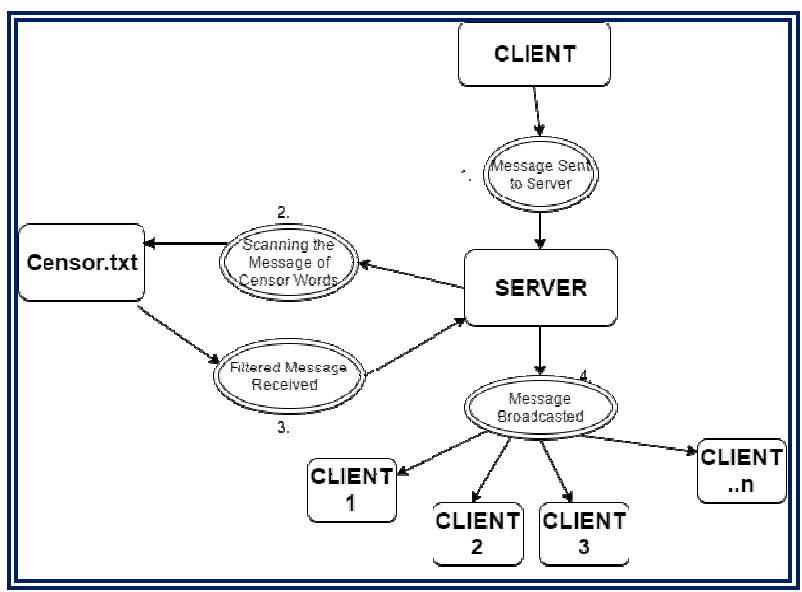

**Figure 2.** Message Processing Using Tool

After the filtering stage of the message, it is then returned back to the server. The filtered message is then broadcasted to one or more clients. This model can be applied on social conversation where users communicate with each other, create groups and freely involve with each other. This model can be applied for Business Communication as well. To understand this better, the code of the algorithm at the client side has been provided below. The tool, based on this model has been tested on Windows Operating System and developed in Java, which is platform independent, hence it would execute successfully on the other platforms.

As soon as the client hits the "Send" button, an Action event is generated which calls the function "actionPerformed (ActionEvent a)" mentioned in the code (Figure 3). A variable, "flag" is set to one to indicate that no flaming has taken place. If the message is "stop", then the connection is automatically terminated and the client is logged off. Otherwise using the "list.iterator()" function, each word in the list of censored words stored in the text file(censor.txt) is scanned in the whole message. For each client, a "count" variable is maintained to indicate the flaming level. As soon as the flame word is encountered, the value of "count" variable is incremented and the "flag" variable is set to zero. After the list is scanned, if the value of "flag" is zero, then a warning message is generated as well as the value of "count" variable is also checked.

```java
@Override
public void actionPerformed(ActionEvent e)
{
    try{
        flag=1;
        DataOutputStream dout=new DataOutputStream(sk.getOutputStream());

    if ((e.getSource() == Send) && (NewMsg.getText() != "")) {
    String str=NewMsg.getText();

        if(str.equalsIgnoreCase("stop"))
        {
        dout.writeUTF("logout#"+name);
        dout.flush();

        }

        else
        {
            Iterator itr=list.iterator();
            while(itr.hasNext())
            {
                String st=(String)itr.next();

                if(str.toLowerCase().indexOf(st.toLowerCase())!= (-1))
                {
                    System.out.println(str.toLowerCase().indexOf(st.toLowerCase()));
                    flag=0;
                    System.out.println("SWEAR WORD FOUND");
                    count++;
                    System.out.println("FLAMING LEVEL:- "+count);
                }
            }
        }
    if(flag==0)
    {
    JOptionPane.showMessageDialog(panel,"Sorry can't send the censor words","Inane warning",
    JOptionPane.WARNING_MESSAGE);
    }
        if(count>=7)  //user is blocked now
            {
                dout.writeUTF("logout#"+name);
        dout.flush();
        JOptionPane.showMessageDialog(panel,"Sorry! You have been blocked","Inane warning",
        JOptionPane.WARNING_MESSAGE);

            }//end

        else if(count<=6)  //if flaming level is low

        {   dout.writeUTF("message#"+name+": "+str+"\n");
        dout.flush();}   //end

        }
}

NewMsg.setText("");
    }

    catch(Exception ex)
    {
        ex.printStackTrace();
    }
}
```

**Figure 3**. Algorithm of the code

If the value of "count" is greater than or equal to seven, then a warning message is displayed and the user is blocked from further communication. If the value of "count" is smaller than or equal to six, then the message is broadcasted to other clients.

### 4.1. WORKING OF THE ALGORITHM

In this section using figure (4-11) we have provided the screenshots of the actual algorithm working in the background of Security Detection Model. There are three people involved in this testing process i.e. Server Administrator, Client 1, Client 2.

**Figure 4**. Client 1

**Figure 5.** Client 2

**Figure 6**. Server Administrator

The first and second windows ask the respective clients their Username, while the third window which belongs to the server confirms the connection between the clients. Both the clients have been authenticated and can exchange messages.

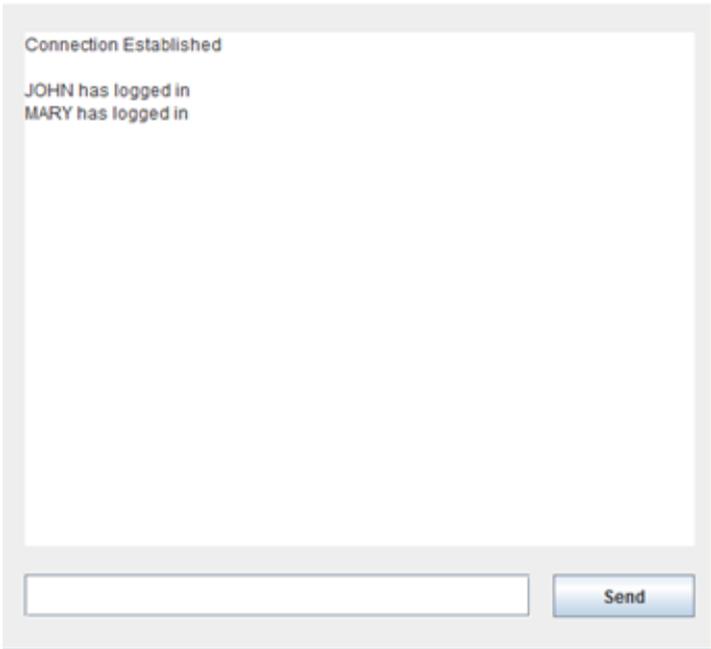

**Figure 7.** Chat Window 1

After a successful connection, the above window appears for both the clients where they can enter their message in the text area adjacent to the "Send" button.

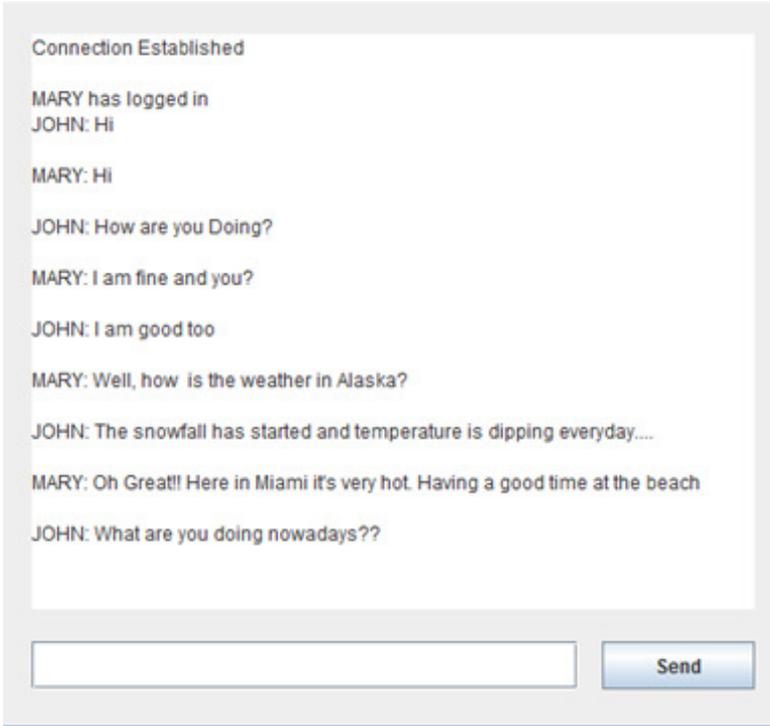

**Figure 8.** Chat Window 2

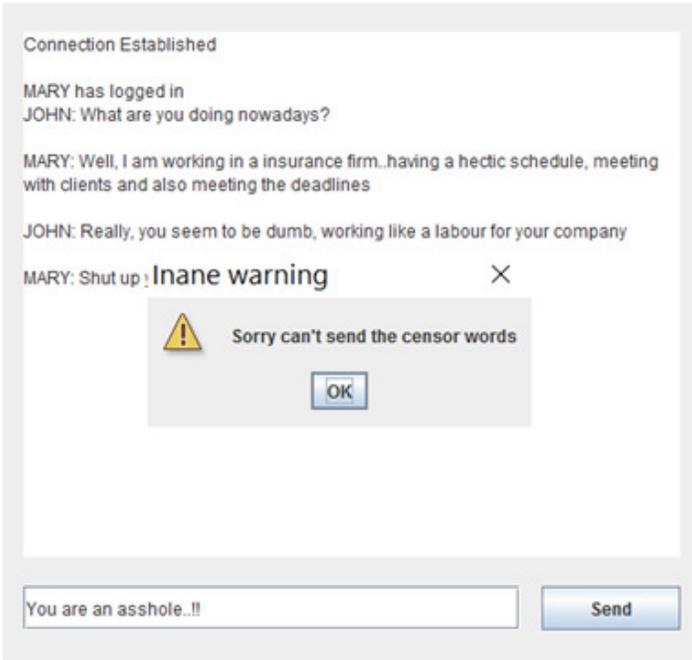

**Figure 9.** Flame Word Sent

In figure 9, the client i.e. MARY tries to send a flame word, "asshole" but a warning window appears which notifies the client that the word cannot be sent and is instead replaced by an asterisk(*).

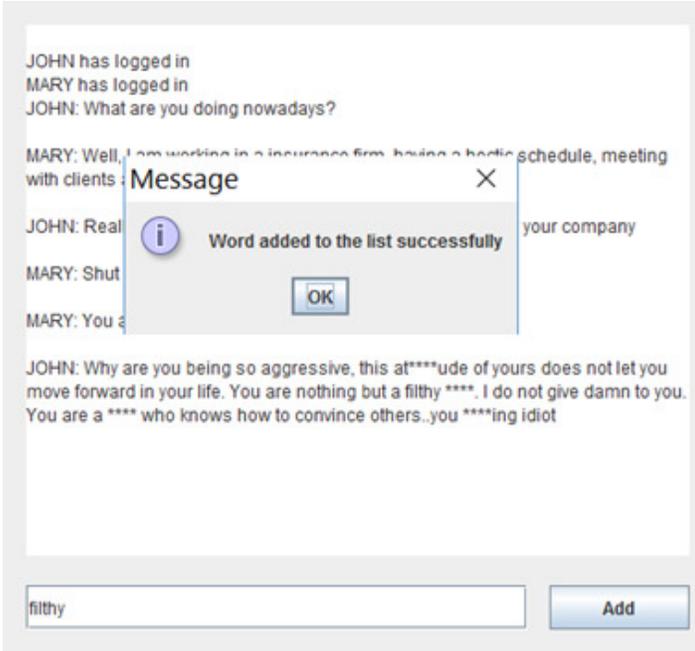

**Figure 10**. Word Added by Administrator

In figure 10, in the last message sent by Client i.e. John, the word "filthy" has not been censored which is then appended to the list by the administrator and a confirmation message is received.

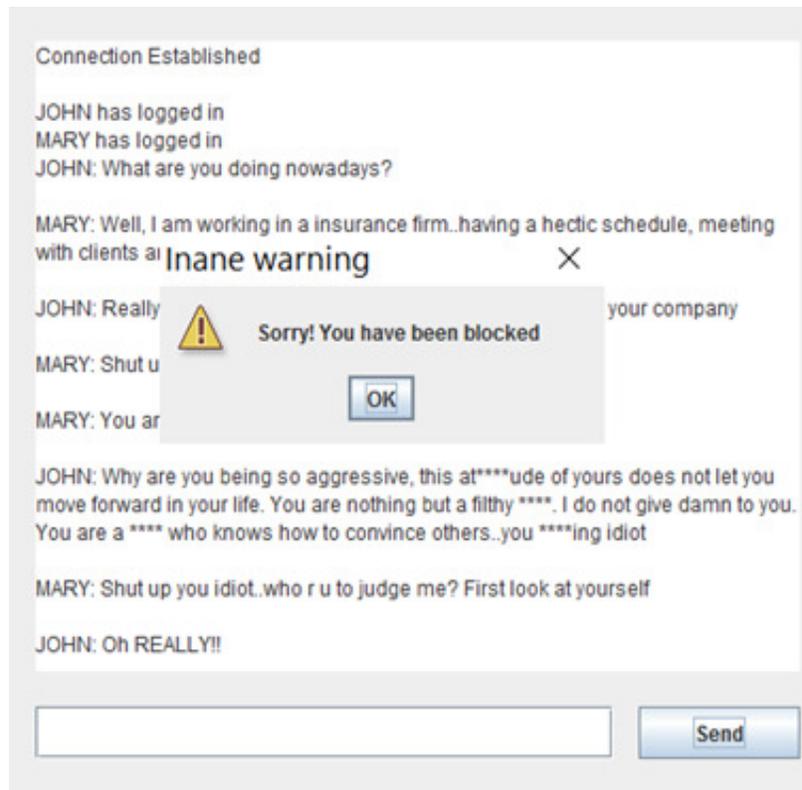

**Figure 11**. Blocking of User

In figure 11, as soon as the flaming level of a client reaches above the threshold value, the above warning message is received and the client is blocked by automatically logging them out.

## 4.2. EXPERIMENTAL RESULTS

The above tested client-to-client communication shows how the user is notified beforehand of censor words being sent to the other user. It also shows that the server administrator has the rights to be able to append more words in the censor list, if they are not included. Further, it has also been depicted that a user is blocked from an ongoing conversation when the flaming level reaches beyond the threshold value.

## 4.3. LIMITATIONS OF SECURITY DETECTION MODEL

There are many censor words which need to be regularly updated in the thesaurus (*censor.txt*) and this cannot be done manually by an administrator since it is a cumbersome job. Therefore, some kind of mechanism like an *administrator bot* needs to be devised which is able to detect new censor words and accommodate them in the database.

## 5. CONCLUSION

The convenience and advantages of online communication is evident to one and all. The extensive rights granted allows a user to share information by posting blogs and exchanging messages through the medium of chatting. However, these rights have eventually led to the hostile and aggressive exchange of words. To keep the immense utility of online communication intact, it is essential to prevent it from getting adulterated with aggressive and abusive form of behaviour. This paper presents the working of a Security Detection tool based on security detection model devised to mark the presence of flaming via a *flame detector tool* and prohibit the user from further social and business conversation. The paper demonstrated that our system is working and is achieving the intended goals.

## Authors

**Shubhankar Gupta:** Currently, a student of B. Tech., 4th Year (Final) Computer Science and Engineering, JIIT, Noida, India. His areas of research interest includes Social Networks especially Cyber Security and Flaming, Database Management using NoSQL technologies, and ERP on the Salesforce.com platform .

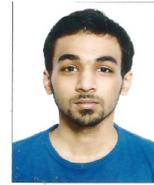

**Nitin:** Professor Nitin is an IBM certified engineer and Senior Member-IEEE and IACSIT, Life Member of IAENG, and Member-SIAM and ACIS. He has more than 155 research papers in peer reviewed International Journals and Transactions, Book Chapters, Symposium, Conferences and Position. His research interest includes Social Networks especially Computer Mediated Communications and Flaming, Interconnection Networks and Architecture, Fault-tolerance and Reliability, Networks-on-Chip, Systems-on-Chip, and Networks-in-Packages, Wireless Sensor Networks. Till now he has guided 08 PhDs, 18 M.Tech. Theses and 60 B.Tech. Major Projects.

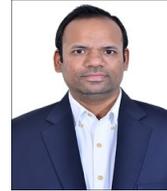

Recently, he has moved to College of Engineering and Applied Sciences, University of Cincinnati, Cincinnati, USA from Jaypee of Institute of Information Technology, Noida, India. He is Associate Editor of Journal of Parallel, Emergent and Distributed Systems, Taylor and Francis, UK. He is referee for the Journal of Parallel and Distributed Computing, Computer Communications, Computers and Electrical Engineering, Mathematical and Computer Modelling, Elsevier Sciences.